\documentclass[letterpaper, 10 pt, conference]{ieeeconf}  

\usepackage{graphics} 
\usepackage{epsfig} 
\usepackage{times} 
\usepackage{amsmath} 
\usepackage{amssymb}  
\usepackage{color}
\usepackage{comment}
\usepackage{bm}
\usepackage{algorithm}
\usepackage{algpseudocode}

\usepackage{subcaption}

\usepackage{hyperref}

\title{\LARGE \bf
Risk-Aware Hosting Capacity Analysis for Flexible Load Interconnection in Distribution Networks
}

\author{
Gobinda Chandra Sarker and Nathan Dahlin
}

\begin{document}

\maketitle
\thispagestyle{empty}
\pagestyle{empty}

\begin{abstract}
The increasing penetration of flexible loads, such as electric vehicles and AI data-centers necessitates new methodologies for quantifying electrical load hosting capacity under operational constraints and flexible connection agreements. We propose a risk-aware hosting capacity framework that explicitly accounts for both flexibility, in the form of load curtailment, and system reliability. The proposed method incorporates a Conditional Value-at-Risk (CVaR) constraint to control the tail risk of excessive curtailment, ensuring that extreme interventions remain limited. Additionally, a weighted $\ell_1$ approach is introduced to limit the number of utility-controlled interventions, enabling control over the frequency of curtailment actions. A regularization parameter is used to tune the intervention count to a desired intervention budget. The resulting optimization formulation is convex and efficiently solvable, allowing scalable implementation. Numerical results demonstrate that the proposed method significantly increases hosting capacity while maintaining strict risk guarantees and limiting intervention frequency, providing a practical balance between flexibility and reliability in distribution systems. The code and dataset for this study is openly available at \href{https://github.com/gcsarker/flexible_load_hosting}{\texttt{github.com/gcsarker/flexible\_load\_hosting}}
\end{abstract}


\section{INTRODUCTION}
Flexible loads refer to electricity demands that do not require an immediate power supply and can be served within a permissible time window \cite{crozier2025potential, norris2025rethinking}. This temporal flexibility allows system operators to shift consumption or curtail load when generation or network capacity is limited. Electricity demand in modern power systems is growing rapidly. According to the International Energy Agency (IEA), global electricity demand is projected to grow by 3.6\% annually over the next five years, compared to 2.8\% annually in the past decade \cite{iea2026electricity}. A significant portion of this growth is driven by emerging power-intensive applications that inherently possess flexibility, such as scheduling artificial intelligence (AI) workloads in data centers, charging electric vehicles (EVs) during off-peak hours, and adaptive HVAC operation in buildings \cite{bogmans2025power}. In particular, data centers alone are expected to contribute up to 50\% of electricity demand growth by 2030 \cite{iea2026electricity}. Effectively coordinating these flexible loads therefore presents a critical opportunity to improve grid utilization, defer infrastructure upgrades, and maintain system reliability.

Recent studies further highlight the practical potential of flexible loads. For example, \cite{radovanovic2022carbon} showed that delay-tolerant data center workloads can be scheduled based on carbon intensity forecasts, enabling significant emissions reductions without affecting service quality, while \cite{yuan2025investigation} demonstrated that HVAC systems in grid-interactive buildings can provide substantial demand-side flexibility through load shifting and adaptive scheduling. Similarly, \cite{almutairi2024hierarchical} proposed a two-layer optimization framework that jointly considers demand response (DR), transmission expansion planning, and hosting capacity to enable higher penetration of EVs and renewables. They showed that EV charging flexibility can significantly increase grid utilization by redistributing demand away from peak periods.

Despite these advances, expanding grid infrastructure at the same pace as demand growth remains economically and operationally challenging. However, apart from the peak hour most time periods, the grid is available to supply to additional load demand which gives us a valid window to tap into for flexible loads. Recent work suggests that up to 76 GW of additional load could be integrated into the U.S. grid without increasing generation capacity, provided that limited and controlled interventions are allowed \cite{norris2025rethinking}. These findings motivate the need for systematic methods to quantify and utilize such flexibility.

DR has been widely studied as a mechanism to enhance grid flexibility by adjusting consumption in response to price signals or system conditions \cite{almutairi2024hierarchical, ahmad2025demand}. Traditional DR programs primarily focus on short-term peak shaving or load shifting, serving as an alternative to costly infrastructure expansion \cite{norris2025rethinking, yuan2025investigation}. However, these approaches typically do not provide explicit guarantees on the frequency and severity of interventions, nor do they directly address the problem of maximizing connectable flexible load under reliability constraints \cite{yasmin2024survey, 10577755}. As a result, they are not well suited for planning-scale integration of large, delay-tolerant loads. More recently, \cite{gu2025role} shows that significant hosting capacity can be unlocked through infrequent interventions under a flexible connection paradigm. However, their formulation relies on a combinatorial, order-statistics-based characterization of intervention events, resulting in a nonconvex problem structure that limits scalability and analytical tractability.

\begin{table}[t]
\centering
\caption{Notations}
\begin{tabular}{lcc}
\hline
\textbf{Definition} & \textbf{Notation} \\
\hline
Hosting capacity & $P$ \\
Feeder Capacity & $\bar{p}_0$ \\
Aggregated load at bus & $\mathcal{L}(t)$ \\
Residual feeder capacity & $\Delta P(t)$ \\
Normalized load profile & $\hat{l}(t)$ \\
New Load at bus & $\hat{P}$ \\

Curtailment amount & $p_{\text{curt}}(t)$ \\
Curtailment depth limit & $\rho$ \\
\hline
\end{tabular}
\label{notation table}
\end{table}

In contrast, this study proposes a convex, risk-aware framework for hosting capacity maximization that explicitly models both the frequency and severity of curtailment. By leveraging a Conditional Value-at-Risk (CVaR)-based constraint, the proposed approach provides a tractable and scalable formulation that guarantees bounded load curtailment even under rare but high-impact stress scenarios. Specifically, we address the following research question: \textbf{how can flexible loads be integrated into distribution networks while ensuring that curtailments remain sparse and bounded, even during rare stress events?} To this end, we develop a stochastic optimization framework that evaluates system performance over a planning horizon and enforces risk-aware constraints on curtailment behavior. The proposed model ensures that operational limits, such as voltage and line constraints, are satisfied all the time while maximizing the amount of connectable load.

The remainder of this paper is organized as follows. Section \ref{formulation} presents the problem formulation, including the curtailment flexibility model in \ref{curtailment_flexibility_model}, power flow constraints in \ref{power_flow}, and the proposed risk-aware hosting capacity framework in \ref{risk_aware}. Section \ref{empirical observation} provides numerical experiments demonstrating the effectiveness of the approach. Section \ref{conclusion} concludes the paper and discusses future research directions. The notation used throughout the paper is summarized in Table~\ref{notation table}.

\section{FORMULATION} \label{formulation}
We consider a single-phase radial distribution feeder with bus set $\mathcal N_0 := \{0\}\cup \mathcal N$, where bus \(0\) denotes the root (substation) bus and \(\mathcal N=\{1,\dots,n\}\) is the set of downstream buses. We assume that a sufficiently rich historical dataset of power demand is available, which enables accurate characterization of load behavior over time. In practice, such a dataset may be used to construct load forecasts before estimating hosting capacity. However, the focus of this paper is to estimate the hosting capacity directly from historical demand data. Let the planning horizon be denoted by $\mathcal{T}$, which consists of discrete time intervals indexed by $t \in \mathcal{T}$. For U.S. distribution utilities, each interval typically corresponds to a 15-minute period \cite{gu2025role}. Our goal is to determine the maximum capacity \(P\) of a new flexible load, such as an EV charging station or an AI data center, connected at bus \(i^\star\in\mathcal N\), under a flexible connection scheme while satisfying all network constraints. Given a normalized load profile $0\leq \hat{l}(t)\leq 1$ representing the temporal variation of the new load, the additional demand at time interval $t$ can be expressed with $P\hat{l}(t)$.

\subsection{Curtailment Flexibility Model} \label{curtailment_flexibility_model}

To maximize the hosting capacity of the distribution feeder, we consider a flexible connection scheme in which during grid stress periods, a portion of the newly connected load can be curtailed 
for some proportion of time agreed to beforehand by the client. Let $p_{\text{curt}}(t) \ge 0$ denote the curtailed portion of the flexible load at time interval $t \in \mathcal{T}$. The actual served load is therefore given by

\begin{equation}\nonumber 
    \hat{P}(t) = P \hat{l}(t) - p_{\text{curt}}(t).
\end{equation}

To ensure that curtailment magnitude remains bounded, we impose a \emph{curtailment depth constraint}:
\begin{equation}
    0 \le p_{\text{curt}}(t) \le \rho \, P\hat{l}(t), \quad \forall t \in \mathcal{T},
    \label{curtailment_depth_bound}
\end{equation}
where $\rho \in [0,1]$ represents the maximum allowable fraction of curtailed load. 

\subsection{Power-Flow Constraint} \label{power_flow}
In this study, we consider a single phase radial distribution network, which is modeled by linearized DistFlow (LinDistFlow) equations. The radial network is represented by a directed graph $(\mathcal{N}, \mathcal{E})$, where $\mathcal{N}$ denotes the set of buses and $\mathcal{E}$ denotes the set of distribution lines. For each line $(i,j) \in \mathcal{E}$, let $r_{ij}$ and $x_{ij}$ denote the resistance and reactance, respectively.

Let \(P_{ij}(t)\) and \(Q_{ij}(t)\) denote the active and reactive line flows from bus \(i\) to bus \(j\) at time \(t\), and let \(V_i(t)\) denote the squared voltage magnitude at bus \(i\). Let \(\mathbf p(t)\in\mathbb R^n\) and \(\mathbf q(t)\in\mathbb R^n\) denote the active and reactive bus load vectors over buses in \(\mathcal N\). Under LinDistFlow,
\begin{equation}\nonumber
P_{ij}(t) = \sum_{k:(j,k)\in \mathcal{E}} P_{jk}(t) + p_j(t), \quad \forall (i,j)\in\mathcal{E},
\end{equation}
\begin{equation}\nonumber
Q_{ij}(t) = \sum_{k:(j,k)\in \mathcal{E}} Q_{jk}(t) + q_j(t), \quad \forall (i,j)\in\mathcal{E}.
\end{equation}
\begin{equation}
V_j(t) = V_i(t) - 2\big(r_{ij} P_{ij}(t) + x_{ij} Q_{ij}(t)\big), \quad \forall (i,j)\in\mathcal{E}.
\label{voltage_drop_eqn}
\end{equation}
The operational constraints are
\begin{align}
    \nonumber v_i^{\min} \le V_i(t) \le v_i^{\max}, 
    \qquad &\forall i\in\mathcal N,\ \forall t\in\mathcal T, \\
    \label{eq:feeder_limit}
    \sum_{i \in \mathcal N} P_i(t) \le \bar P_0, 
    \qquad &\forall t\in\mathcal T.
\end{align}

Let \(\mathcal{L}\in\mathbb R^n\)  denote the aggregated baseline load vector. If the new flexible load is connected at bus \(i^\star\), then the total active load vector becomes
\begin{equation}
    \mathbf p(t)= \mathcal{L} (t)+\big(P\hat{l}(t)-p_{\text{curt}}(t)\big)\mathbf e_{i^\star},
    \label{p_total}
\end{equation}
where \(\mathbf e_{i^\star}\in\mathbb R^n\) is the \(i^\star\)-th elementary vector with a $1$ at entry $i^\star$ and zeros elsewhere. Equation \eqref{voltage_drop_eqn} can be written in matrix form in terms of resistance and reactance matrices 
$\textbf{R, X} \in \mathbb{R}^{n\times n}$ along with the real and reactive load injection $\textbf{p}(t)$ and $\textbf{q}(t)$ respectively as 
\begin{equation}
    \mathbf V(t)=v_0\mathbf 1-2\mathbf R\,\mathbf p(t)-2\mathbf X\,\mathbf q(t),
    \label{voltage_matrix}
\end{equation}
Where $v_0 = v_{max}$ is the squared voltage magnitude at the substation. Given load power factor and the corresponding ratio of reactive to real power $\eta$, we can write, \(\mathbf q(t)=\mathrm{diag}(\boldsymbol\eta(t))\mathbf p(t)\). Thus \eqref{voltage_matrix} becomes
\begin{equation}
    \mathbf V(t)=v_0\mathbf 1-2\mathbf Z(t)\mathbf p(t),
    \label{voltage_z_form}
\end{equation}
where
$$
\mathbf Z(t):=\mathbf R+\mathbf X\,\mathrm{diag}(\boldsymbol\eta(t)).
$$
Enforcing the lower-voltage constraint \(V_j(t)\ge V_j^{\min}\) for each bus \(j\in\mathcal N\), and substituting \eqref{p_total} into \eqref{voltage_z_form}, yields
\begin{equation}
    P\hat{l}(t)-p_{\text{curt}}(t)
    \le
    \frac{v_0-v_j^{\min}-2\mathbf Z_j(t)\mathcal{L}(t)}
    {2Z_{j i^\star}(t)},
    \qquad \forall j\in\mathcal N,
    \label{voltage_capacity_eachbus}
\end{equation}
where \(\mathbf Z_j(t)\) denotes the \(j\)-th row of \(\mathbf Z(t)\). In addition, the substation feeder power is upper bounded by $\bar{p}_0$, which satisfies 
\begin{equation}
   P\hat{l}(t) - p_{\text{curt}}(t) \le \bar{p}_0- \mathcal{L}(t)
    \label{transformer_capacity}
\end{equation} 

Combining \eqref{voltage_capacity_eachbus} and \eqref{transformer_capacity}, the served flexible load is bounded by the residual feeder capacity
\begin{equation}
    P\hat{l}(t)-p_{\text{curt}}(t)\le \Delta P(t),
    \label{served_residual}
\end{equation}
where
\begin{equation}
    \Delta P(t):=
    \min\left\{
    \bar p_0-\mathcal{L}(t),\
    \min_{j\in\mathcal N}
    \frac{v_0-v_j^{\min}-2\mathbf Z_j(t)\mathcal{L}(t)}
    {2Z_{j i^\star}(t)}
    \right\}.
    \label{residual_capacity}
\end{equation}

\begin{figure}[b]
    \centering
    \includegraphics[width=0.98\linewidth, height = 0.6\linewidth]{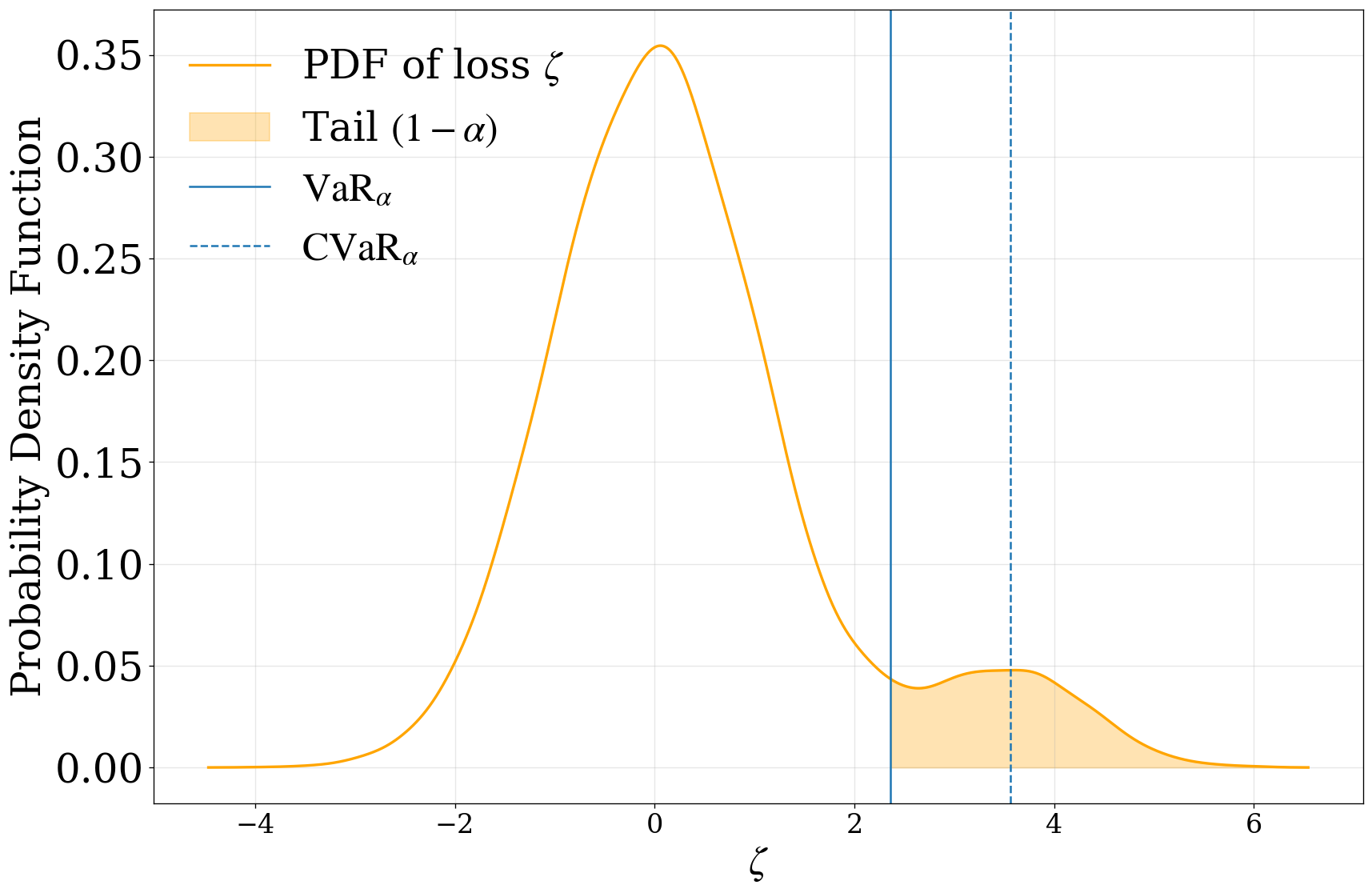}
    \caption{Example PDF of the random loss $\zeta$.  $\text{CVaR}_{\alpha}$ denotes the expected loss within the tail (shaded) region $(1-\alpha)$.}
    \label{example_cvar}
\end{figure}

\subsection{Risk-Aware Hosting Capacity} \label{risk_aware}
We model the risk of large curtailment events exceeding the allowed limit as
\begin{equation}
    \zeta(t) = p_{\text{curt}}(t) - \rho P \hat{l}(t),
\end{equation}
where $\rho \in [0, 1]$ is the nominal fraction of the flexible load demand. We impose a CVaR constraint at confidence level $\alpha \in (0,1)$:
\begin{equation}
\mathrm{CVaR}_{\alpha}\left[\zeta(t)\right] \le \epsilon,
    \label{cvar_curt}
\end{equation}
where $\epsilon$ is the allowable risk level. CVaR is a tail-risk measure that quantifies the expected value of a random variable in the worst-case $(1-\alpha)$ fraction of outcomes \cite{madavan2023risk}. This is illustrated in Fig. \ref{example_cvar} with distribution of random loss $\zeta$. $\mathrm{VaR}_\alpha$ corresponds to the $\alpha$-quantile  of the distribution, while $\mathrm{CVaR}_\alpha$ represents the average of the tail region beyond this threshold, capturing the severity of extreme events.

Introducing an auxiliary variable $\gamma$ and slack variables $s(t) \ge 0$, the epigraph form of the CVaR constraint in Eq. \ref{cvar_curt} is
\begin{align}   
    \label{cvar_epi1}
    & p_{\text{curt}}(t) - \rho P \hat{l}(t) - \gamma \leq s(t), 
    \quad \forall t \in \mathcal{T}, \\
    & \gamma + \frac{1}{(1-\alpha)|\mathcal{T}|} 
    \sum_{t \in \mathcal{T}} s(t) \le \epsilon,\label{cvar_epi2}\\
    &s(t) \geq 0,\quad \forall\,t\in \mathcal{T}.\label{cvar_epi3}
\end{align}
The hosting capacity estimate is obtained by solving the following optimization problem 
\begin{equation}
\begin{aligned}
\max_{P,\gamma,s,p_{\text{curt}}}&\quad 
    P\\
    \text{s.t.}&\quad \eqref{served_residual},\,\eqref{cvar_epi1},\,\eqref{cvar_epi3}\, \forall t\in \mathcal{T} \text{ \& }\eqref{cvar_epi2}.
\end{aligned}
\label{cvar_hosting} 
\end{equation}
\subsection{Sparse Intervention Scheduling} 
Although constraint \eqref{cvar_curt} and its equivalent epigraph form \eqref{cvar_epi1}-\eqref{cvar_epi2} limit curtailment depth, such constraints do not generally provide guarantees on the number of interventions. 
While, for a fixed $\alpha$ the probability of interventions exceeding $\text{VaR}_{\alpha}$ is bounded, small interventions may still occur throughout the planning horizon. We introduce sparsity in curtailment schedules using weighted $\ell_1$ regularization
\begin{equation}\nonumber 
    \sum_{t \in \mathcal{T}} w_{\text{norm}}(t)\, p_{\text{curt}}(t),
\end{equation}
\begin{align}
w(t) &:= \Delta P(t), \\
w_{\text{norm}}(t) &:= \frac{w(t)}{\max (w(t))}.
\label{weight_update}
\end{align}
where $w_{\text{norm}}(t) \in (0,1)$ are normalized weights proportional to $\Delta P(t)$, as shown in  \eqref{weight_update}. This is done to discourage the system from large curtailment when the residual capacity is small. The hosting capacity problem \ref{cvar_hosting} objective can then be augmented with a penalty term as follows:
\begin{equation}
\begin{aligned}
\max_{P,\gamma,s,p_{\text{curt}}}&\quad 
    P - \lambda \sum_{t\in \mathcal{T}}w_{\text{norm}}(t)p_{\text{curt}}(t)\\
    \text{s.t.}&\quad \eqref{served_residual},\,\eqref{cvar_epi1},\,\eqref{cvar_epi3}\,\forall t\in \mathcal{T} \text{ \& }\eqref{cvar_epi2}.
\end{aligned}
\label{sparse_cvar_hosting}
\end{equation}

\begin{figure}[b]
    \centering
    \includegraphics[width=0.98\linewidth, height = 0.6\linewidth]{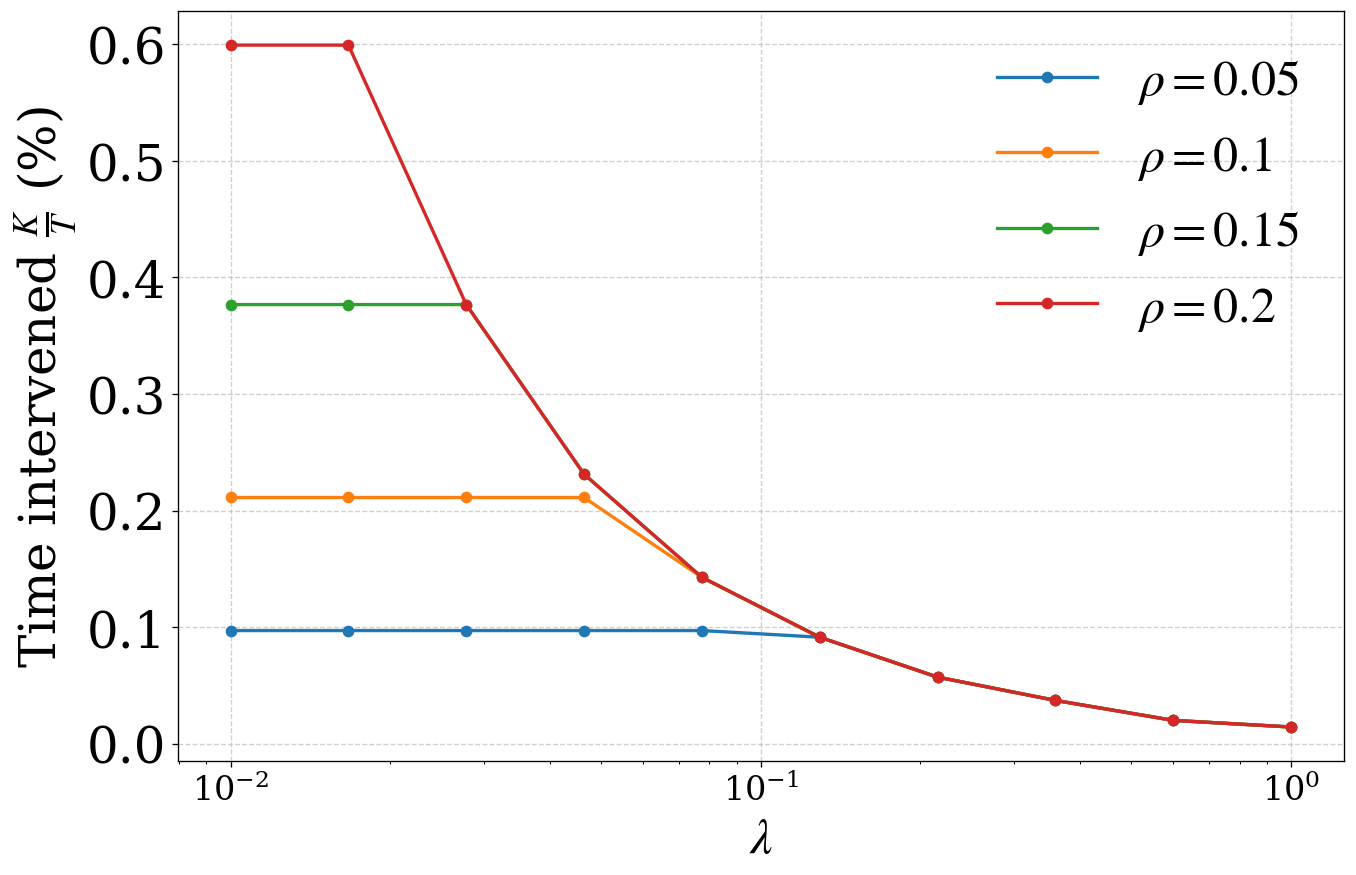}
    \caption{Intervention count with respect to regularization parameter $\lambda$ for different curtailment limits $\rho$.}
    \label{N_int_v_lambda}
\end{figure}

\begin{figure*}[!t]
    \centering

    \begin{subfigure}[t]{0.45\linewidth}
        \centering
        \includegraphics[width=\linewidth, height = 0.6\linewidth]{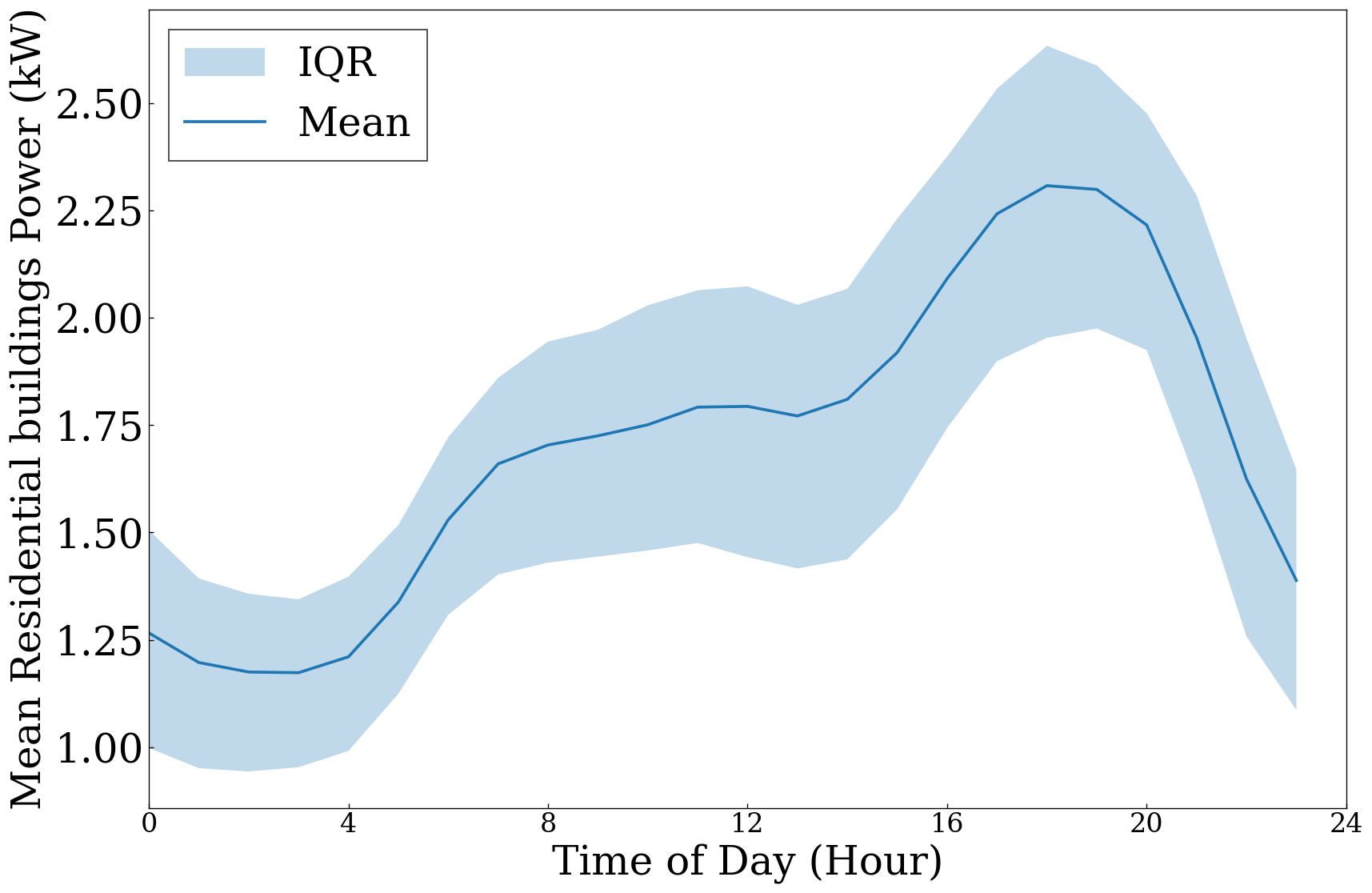}
        \caption{Daily base load profile}
        \label{base_load_shape}
    \end{subfigure}
    \hfill
    \begin{subfigure}[t]{0.45\linewidth}
        \centering
        \includegraphics[width=\linewidth, height = 0.6\linewidth]{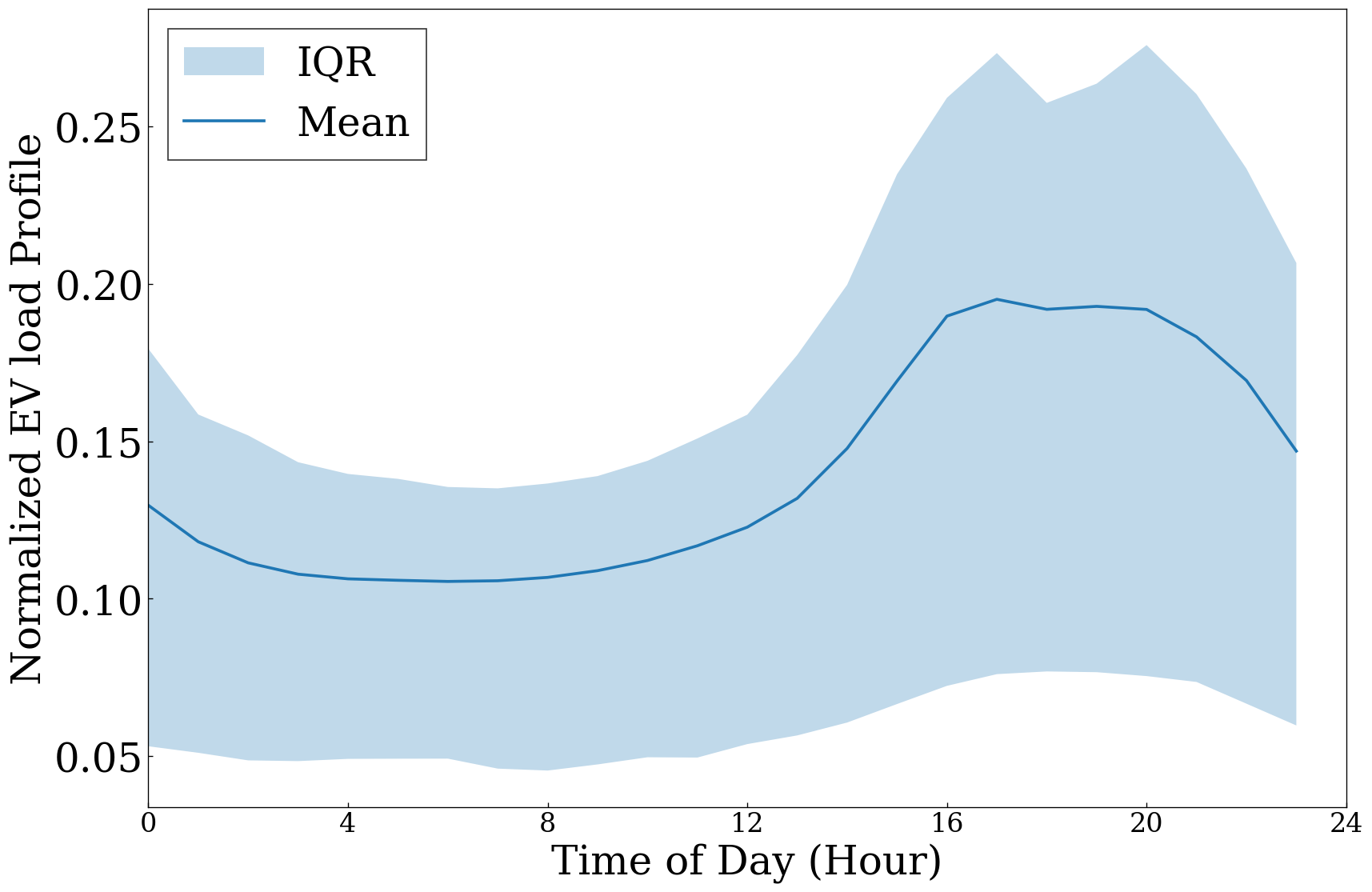}
        \caption{Normalized EV charging profile}
        \label{ev_load_shape}
    \end{subfigure}

    \caption{ Typical average daily profiles of bus loads and EV charging demand, showing mean behavior and variability (IQR) across time of day.}
    \label{load_profiles}
\end{figure*}

Here, $\lambda > 0$ is a regularization parameter that controls the sparsity of curtailment actions. 
Larger values of $\lambda$ discourage frequent interventions, resulting in a smaller number of interventions. 

Define the number of desired intervention periods as
    $\mathcal{N}_{int} := \sum_{t \in \mathcal{T}} \mathbf{1}\{p_{\text{curt}}(t) > 0\}$,
where $\mathbf{1}\{x\}$ denotes the indicator function for event $x$. To ensure that the intervention count is less than or equal to target $\mathcal{N}_{int}$, we tune $\lambda$ via a bisection algorithm iteratively. The bisection method is provided in Algorithm \ref{algo_bisection}, which allows fast convergence to the optimal intervention count. The monotonic assumption that the intervention count is non-increasing in $\lambda$ is empirically verified in Fig. \ref{N_int_v_lambda}.

\addtolength{\textheight}{-3cm}   

\section{NUMERICAL EXPERIMENTS}

We conduct our experiments on the IEEE 123-bus test feeder \cite{schneider2017analytic}, modeled as a single-phase radial distribution network. The baseline demand is constructed using the NREL U.S. residential load dataset \cite{wilson2022end} for New York, which is distributed across 106 buses to simulate spatially distributed consumption. To ensure realistic operating conditions, the load profiles are first scaled such that the minimum bus voltage remains above 0.95~p.u., while the substation (transformer) voltage is fixed at 1.0~p.u. The feeder transformer capacity is then set equal to the resulting peak load demand. To create operational headroom for integrating flexible loads, the baseline demand is further reduced by 10--20\%. This allows controlled evaluation of additional load hosting capability under constrained conditions. The EV charging demand profile is adopted from prior work in \cite{sorensen2021analysis} and used as the normalized flexible load signal in our simulations. Fig~\ref{base_load_shape} and~\ref{ev_load_shape} illustrate daily variations of the baseline and EV load profiles, respectively. We inserted the EV charging profile at bus 13 in our experiment. 

\begin{figure*}[t!]
    \centering

    \begin{minipage}{0.48\linewidth}
        \centering
        \includegraphics[width=\linewidth, height = 0.6\linewidth]{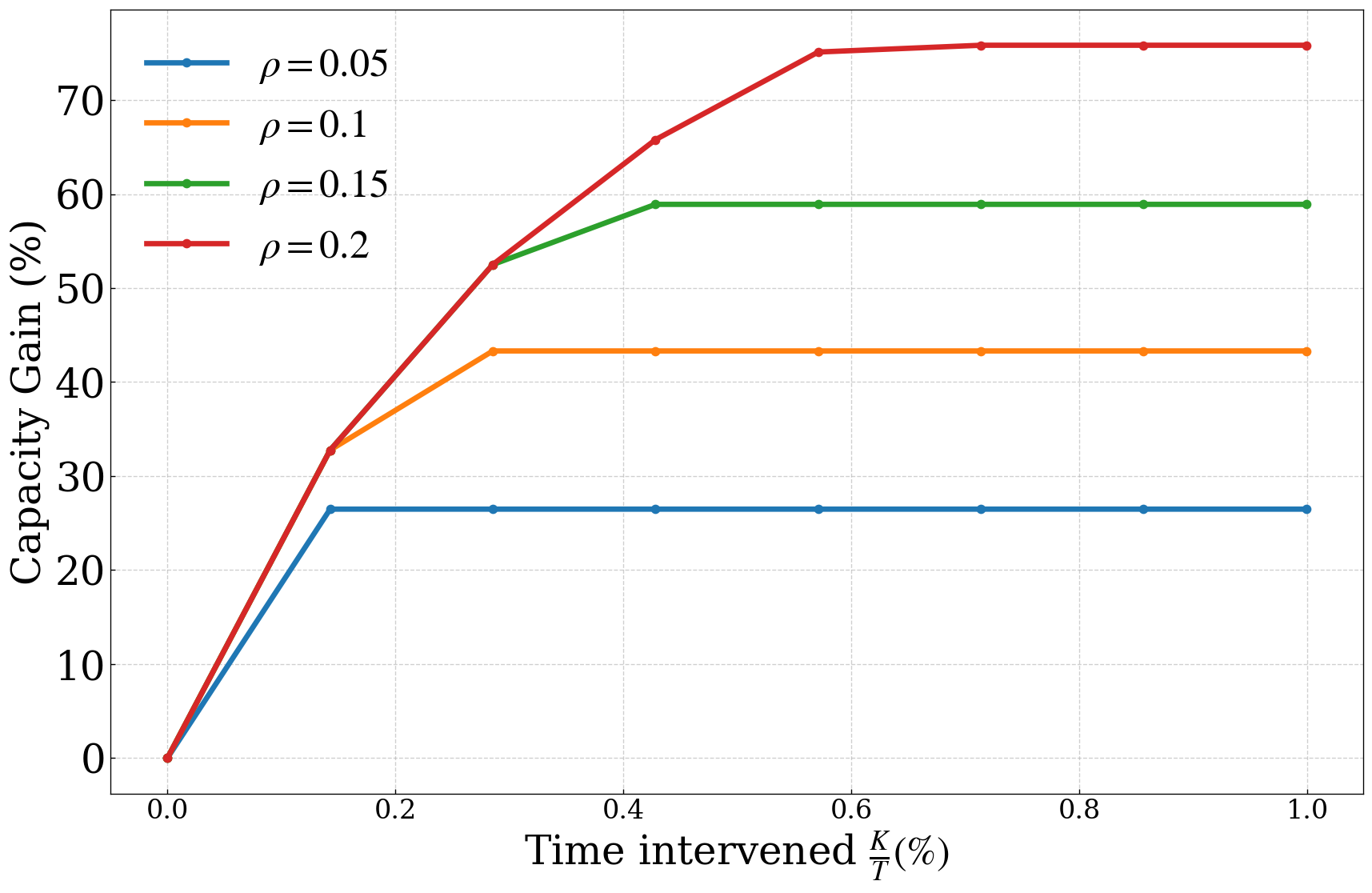}
        \caption{Hosting capacity gain with respect to time intervened (\%) per year.}
        \label{gain_v_k}
    \end{minipage}
    \hfill
    \begin{minipage}{0.48\linewidth}
        \centering
        \includegraphics[width=\linewidth, height = 0.6\linewidth]{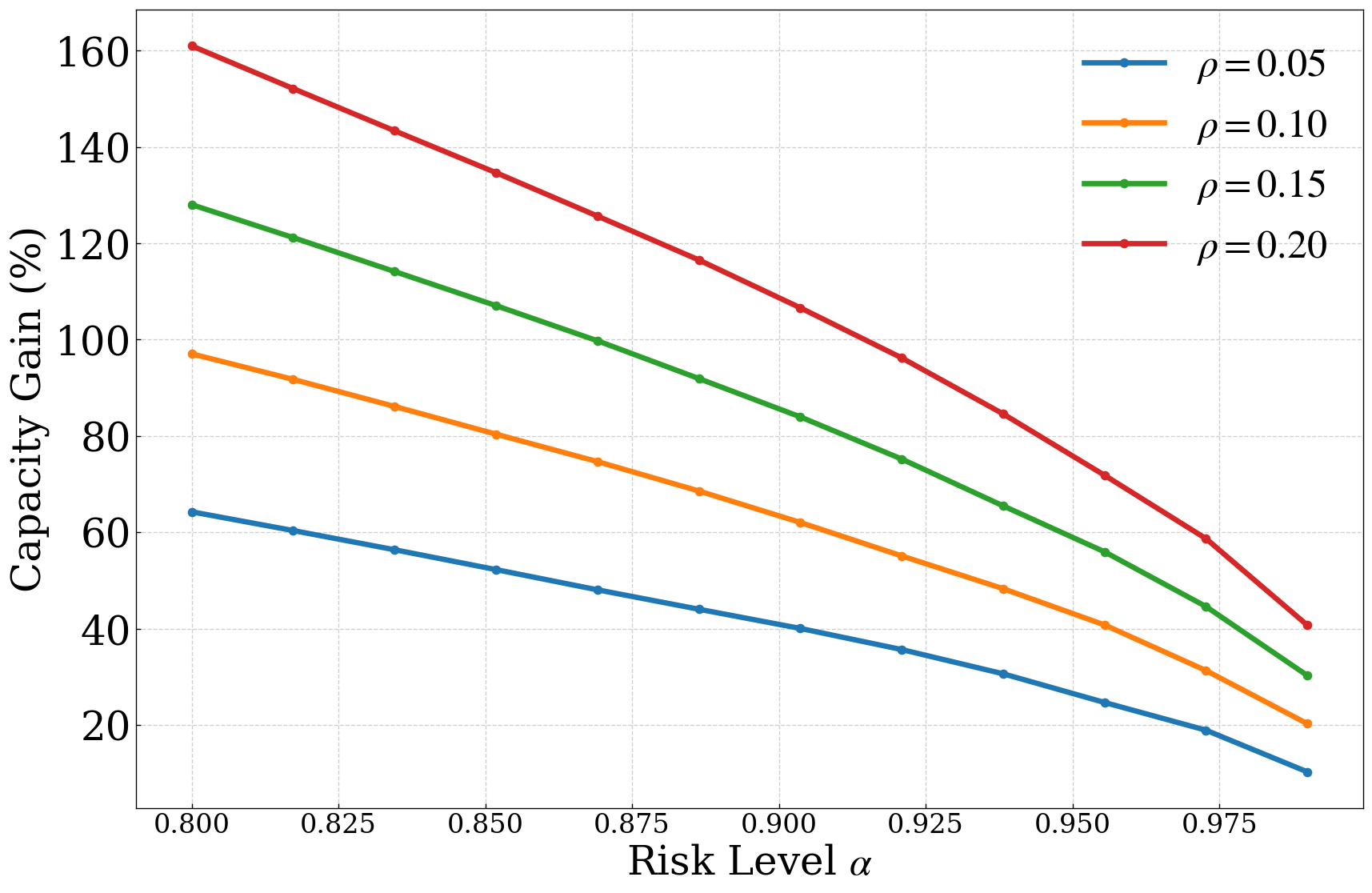}
        \caption{Hosting capacity gain with respect to risk level $\alpha$.}
        \label{gain_v_alpha}
    \end{minipage}

\end{figure*}

\subsection{Empirical Observations} \label{empirical observation}

Fig. \ref{gain_v_k} illustrates the hosting capacity gain as a function of the allowable intervention count $\mathcal{N}_{int}$. When $\mathcal{N}_{int} = 0$, no curtailment is permitted, and the system operates under the inflexible capacity limit. As $\mathcal{N}_{int}$ increases, more intervention opportunities become available, enabling higher levels of flexible load integration and thereby increasing the achievable hosting capacity.

In our experiment, $\mathcal{N}_{int}$ is varied up to $1\%$ the total planning horizon $\mathcal{T}$, reflecting that even small number of interventions unlocks large capacity. Additionally, we examine the impact of the curtailment depth constraint $\rho \in [0, 0.2]$. Larger values of $\rho$ allow deeper curtailment at each intervention, which further enhances the hosting capacity. However, for each fixed $\rho$, the capacity gain eventually saturates as $\mathcal{N}_{int}$ increases. This plateau occurs because the curtailment depth constraint $\rho$ becomes the limiting factor—once the most critical time steps (feeder limit constraint violations) are mitigated, additional interventions do not provide further benefit. Thus, the achievable hosting capacity is jointly constrained by both the intervention budget $\mathcal{N}_{int}$ and the curtailment depth $\rho$.

Fig.~\ref{gain_v_alpha} shows the hosting capacity gain as a function of the risk level $\alpha$ for different curtailment depth limits $\rho$. As $\alpha$ increases, the CVaR constraint becomes more conservative, restricting violations to rarer extreme events and thereby reducing the achievable hosting capacity. For all values of $\rho$, the capacity gain decreases monotonically with $\alpha$. However, larger curtailment depth $\rho$ consistently yields higher capacity gains, since deeper curtailment provides greater flexibility to mitigate critical overload periods.

\begin{figure*}[t]
    \centering

    \begin{subfigure}[t]{0.48\linewidth}
        \centering
        \includegraphics[width=\linewidth, height = 0.7\linewidth]{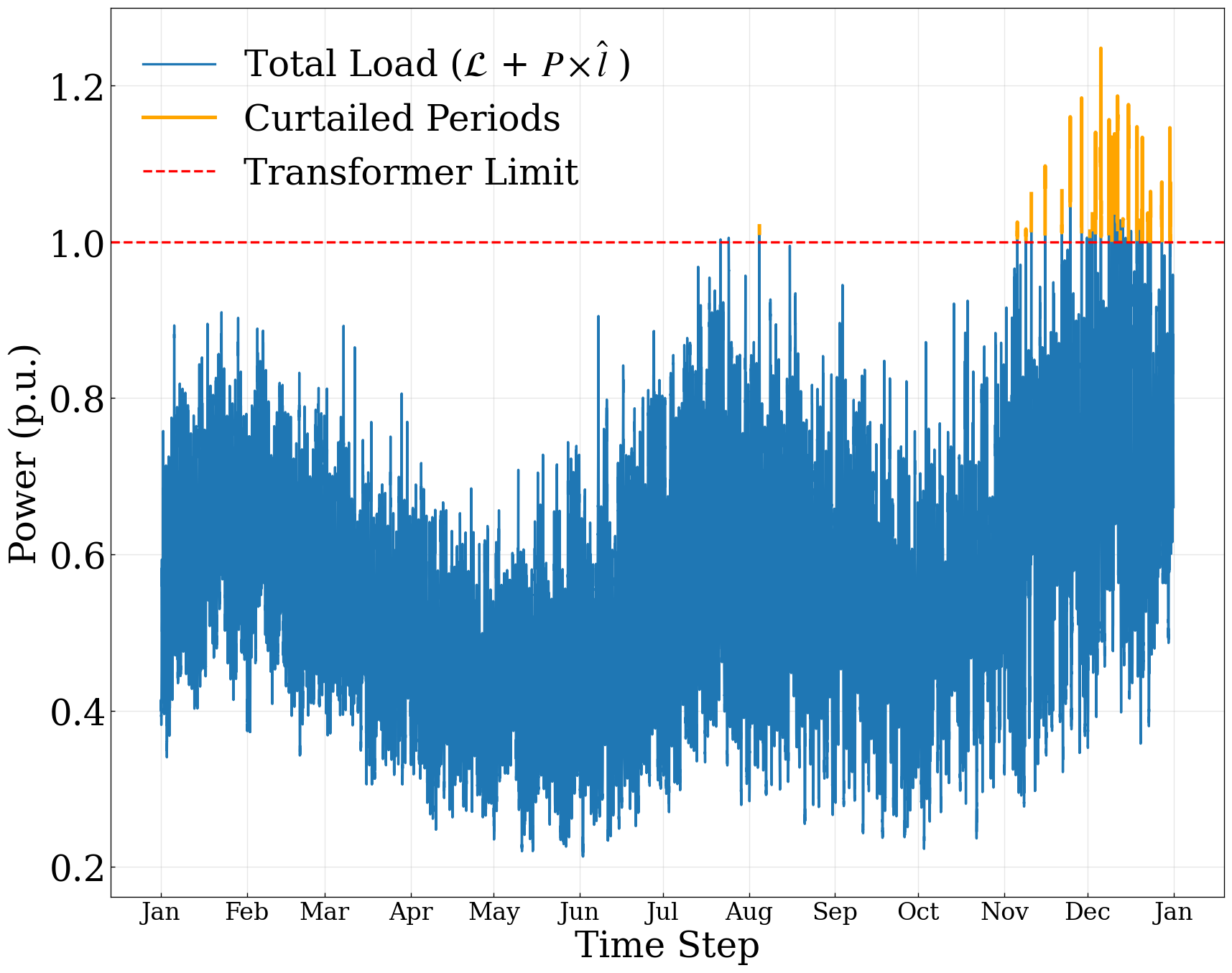}
        \caption{$(\mathcal{L}(t) + P\hat{l}(t))$}
        \label{load_wo_curtailment}
    \end{subfigure}
    \hfill
    \begin{subfigure}[t]{0.48\linewidth}
        \centering
        \includegraphics[width=\linewidth, height = 0.7\linewidth]{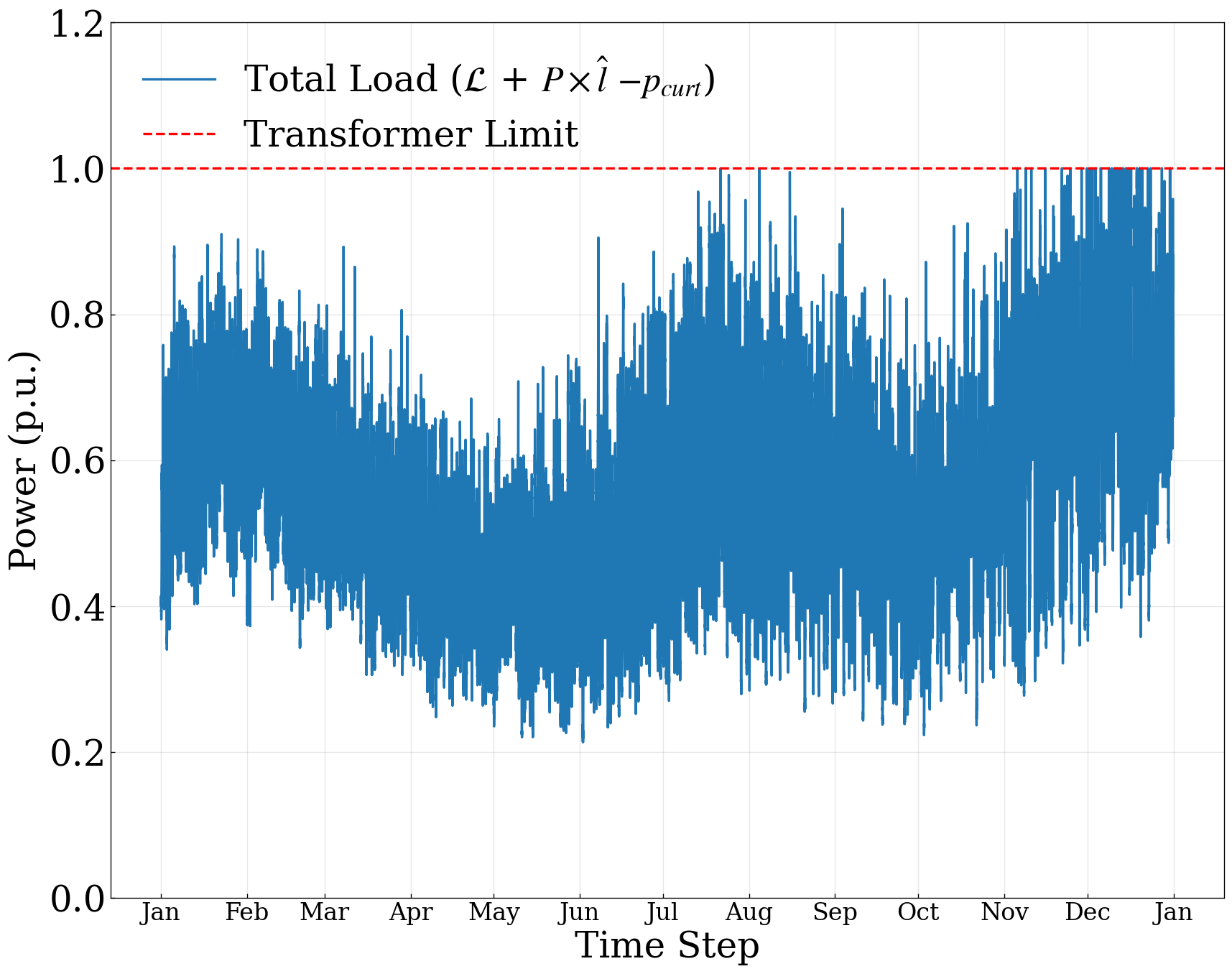}
        \caption{$\mathcal{L}(t) + P\hat{l}(t) - p_{\text{curt}}(t)$}
        \label{load_w_curtailment}
    \end{subfigure}

    \caption{Impact of flexible load hosting and curtailment on feeder capacity (a) Total load $(\mathcal{L}(t) + P\hat{l}(t))$. Highlighted periods shows where curtailment is applied as the load exceeds the transformer limit. (b) Total load profile ($\mathcal{L}(t) + P\hat{l}(t) - p_{\text{curt}}(t)$) after optimal curtailment, satisfying the transformer capacity constraint}
    \label{load_curtailment_viz}
\end{figure*}

Fig.~\ref{load_curtailment_viz} compares the total load profile before and after applying optimal curtailment. In Fig.\ref{load_wo_curtailment}, the total load $\mathcal{L}(t) + P\hat{l}(t)$ exceeds the feeder limit $\bar{P}$ during peak periods, indicating potential overloads. The highlighted regions correspond to the critical time set $\mathcal{T}_{\text{crit}}$, where curtailment is required. In Fig.\ref{load_w_curtailment}, the optimized curtailment $p_{\text{curt}}(t)$ is applied, resulting in the adjusted load $\mathcal{L}(t) + P\hat{l}(t) - p_{\mathrm{curt}}(t)$ remaining within the transformer limit. This demonstrates that targeted curtailment during critical periods effectively mitigates overload violations while preserving the overall load profile.

\begin{figure*}[h]
    \centering

    \begin{subfigure}[t]{0.48\linewidth}
        \centering
        \includegraphics[width=\linewidth, height = 0.6\linewidth]{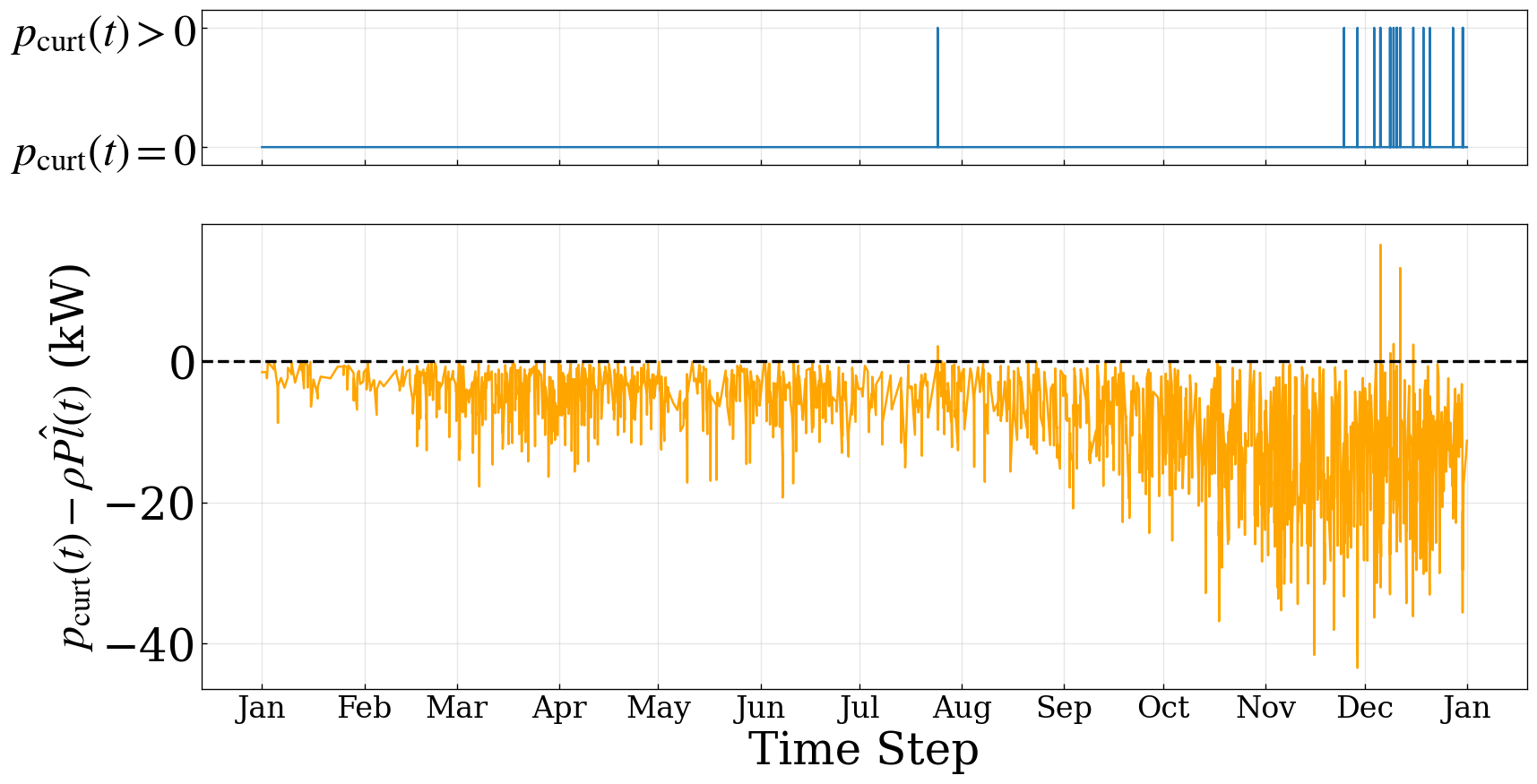}
        \caption{(Top) Sparse intervention schedule. (Bottom) A small fraction of the time-steps where violations $(p_{\text{curt}} > \rho P\hat{l}(t))$ occurred. The curtailment violation is effectively minimized by the CVaR constraint. }
        \label{curtailment_cap_violation}
    \end{subfigure}
    \hfill
    \begin{subfigure}[t]{0.48\linewidth}
        \centering
        \includegraphics[width=\linewidth, height = 0.6\linewidth]{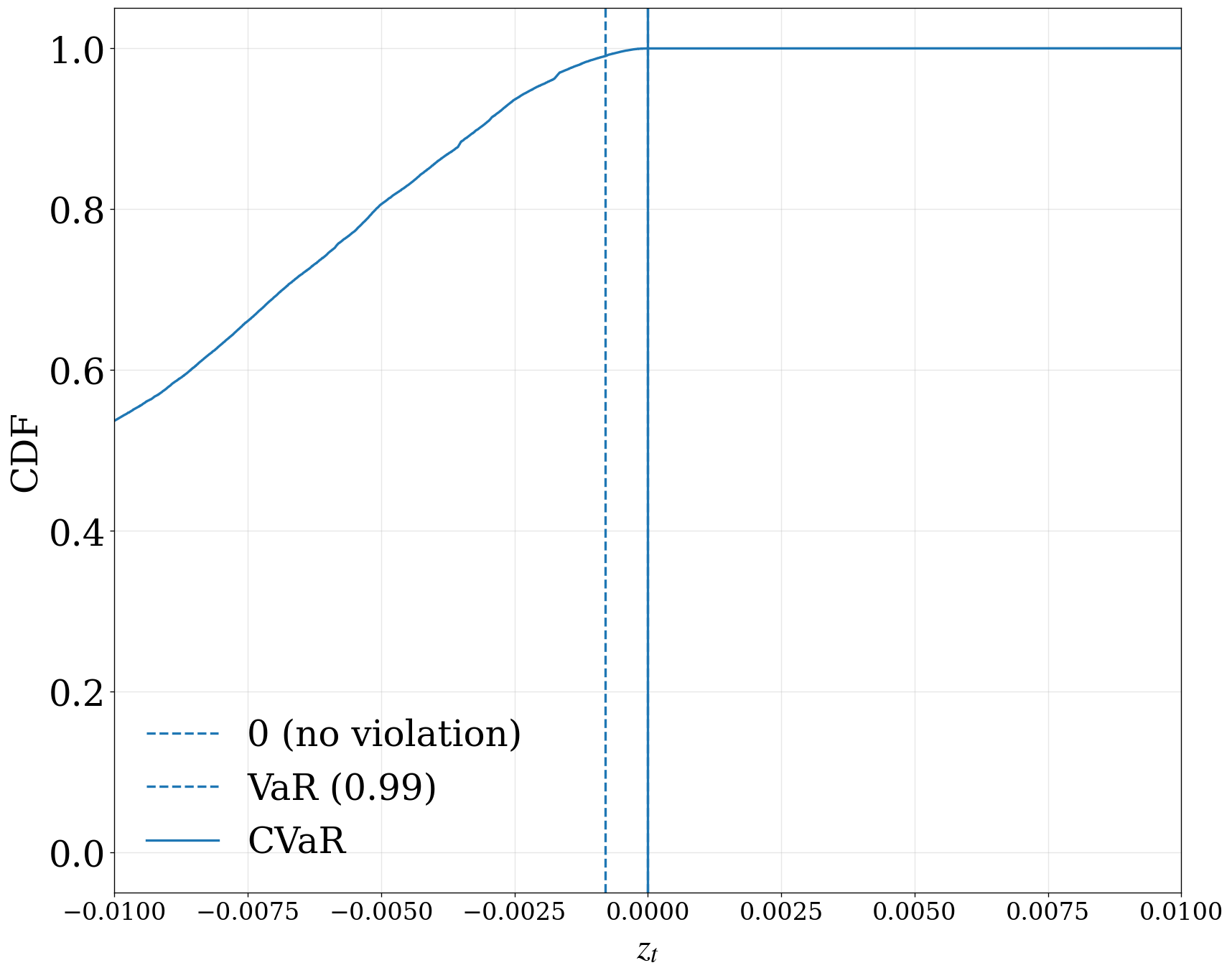}
        \caption{Empirical CDF of the curtailment violation, $\zeta(t) = p_{\text{curt}}(t) - \rho P \hat{l}(t)$. only a few values contribute to the tail $(1-\alpha)$}
        \label{cdf_tail}
    \end{subfigure}

    \caption{Illustration of Curtailment violation and the impact of CVaR constraint}
    \label{cvar_tail}
\end{figure*}

In Fig.~\ref{curtailment_cap_violation}, the applied curtailment $p_{\mathrm{curt}}(t)$ is compared against the depth constraint $\rho P \hat{l}(t)$, showing that the intervention respects the prescribed limit most of the time. This is due to the CVAR relaxation of the curtailment bound. We can also verify from Fig.~\ref{cdf_tail}, which presents the empirical cumulative distribution function (CDF) of the violation metric $\zeta(t)$. The distribution lies predominantly at or below zero, indicating that the curtailment depth violations are effectively minimized. The $\mathrm{VaR}_{\alpha}$ (at $\alpha = 0.99$) is non-positive, meaning that most of the time the curtailment is under the depth constraint. The corresponding $\mathrm{CVaR}_{\alpha}$ remains tightly controlled, verifying that the tail risk constraint is satisfied.

\section{Conclusion} \label{conclusion}

This study proposes a convex, risk-aware framework for hosting capacity analysis of flexible loads in distribution systems. By leveraging a CVaR-based formulation, the approach explicitly controls tail risk associated with excessive curtailment. while $\ell1$ regularization introduces sparsity by limiting the number of intervention periods. Together, these features enable precise and interpretable control over both the magnitude and frequency of flexibility actions. Results demonstrate that substantial increases in hosting capacity can be achieved with only a small number of targeted interventions, revealing that grid constraints are often driven by a few critical time periods. Compared to conventional demand response and prior flexible connection methods, the proposed framework provides stronger reliability guarantees, improved scalability, and direct operator control. A natural extension to our work is the development of a network-wide formulation that jointly optimizes hosting capacity across multiple buses. Future research will also focus on the integration of backup resources such as determining the minimum backup generator capacity to achieve zero curtailment for hosting a target capacity, as well as alternative or complementary forms of load flexibility, including delay tolerance. 



\appendix
\section{Bisection Algorithm} 
\begin{algorithm}
\caption{Bisection tuning of $\lambda$}
\label{algo_bisection}
\begin{algorithmic}[1]
\Require $\Delta P(t)$, $\hat{l}(t)$, target intervention budget $\mathcal{N}_{int}$, bounds $\lambda_{\min},\lambda_{\max}$ and Maximum iterations $N_{max}$.
\For{$k=1,\dots,N_{\max}$}
    \State $\lambda \gets (\lambda_{\min}+\lambda_{\max})/2$
    \State Solve \ref{sparse_cvar_hosting} and obtain $p_{\text{curt}}(t)$
    \State $\mathcal{N}_{\mathrm{current}} \gets \sum_{t=1}^{T}\mathbf{1}\{p_{\mathrm{curt}}(t)>0\}$
    \If{$|\mathcal{N}_{\mathrm{current}}-\mathcal{N}_{int}| \le \varepsilon_{\mathcal{N}}$}
        \State \textbf{break}
    \ElsIf{$\mathcal{N}_{\mathrm{current}}>\mathcal{N}_{int}$}
        \State $\lambda_{\min} \gets \lambda$
    \Else
        \State $\lambda_{\max} \gets \lambda$
    \EndIf
\EndFor
\State \Return $\lambda$
\end{algorithmic}
\end{algorithm}



\bibliographystyle{IEEEtran}
\bibliography{references}

@article{bogmans2025power,
  title={Power hungry: How ai will drive energy demand},
  author={Bogmans, Christian and Gomez-Gonzalez, Patricia and Ganpurev, Ganchimeg and Melina, Giovanni and Pescatori, Andrea and Thube, Sneha},
  journal={IMF Working Papers},
  volume={81},
  number={4},
  pages={2025},
  year={2025},
  publisher={International Monetary Fund}
}

@article{crozier2025potential,
  title={The potential of data center energy demand to provide grid flexibility},
  author={Crozier, Constance and Liska, Matthew},
  journal={Current Sustainable/Renewable Energy Reports},
  volume={12},
  number={1},
  pages={12},
  year={2025},
  publisher={Springer}
}

@article{gu2025role,
  title={The Role of Flexible Connection in Accelerating Load Interconnection in Distribution Networks},
  author={Gu, Nan and Chen, Ge and Qin, Junjie},
  journal={arXiv preprint arXiv:2510.11476},
  year={2025}
}

@article{radovanovic2022carbon,
  title={Carbon-aware computing for datacenters},
  author={Radovanovi{\'c}, Ana and Koningstein, Ross and Schneider, Ian and Chen, Bokan and Duarte, Alexandre and Roy, Binz and Xiao, Diyue and Haridasan, Maya and Hung, Patrick and Care, Nick and others},
  journal={IEEE Transactions on Power Systems},
  volume={38},
  number={2},
  pages={1270--1280},
  year={2022},
  publisher={IEEE}
}

@article{schneider2017analytic,
  title={Analytic considerations and design basis for the IEEE distribution test feeders},
  author={Schneider, Kevin P and Mather, BA and Pal, Bikash Chandra and Ten, C-W and Shirek, Greg J and Zhu, Hao and Fuller, Jason C and Pereira, Jos{\'e} Luiz Rezende and Ochoa, Luis F and de Araujo, Leandro Ramos and others},
  journal={IEEE Transactions on power systems},
  volume={33},
  number={3},
  pages={3181--3188},
  year={2017},
  publisher={IEEE}
}

@techreport{wilson2022end,
  title={End-use load profiles for the US building stock: Methodology and results of model calibration, validation, and uncertainty quantification},
  author={Wilson, Eric JH and Parker, Andrew and Fontanini, Anthony and Present, Elaina and Reyna, Janet L and Adhikari, Rajendra and Bianchi, Carlo and CaraDonna, Christopher and Dahlhausen, Matthew and Kim, Janghyun and others},
  year={2022},
  institution={National Renewable Energy Laboratory (NREL), Golden, CO (United States)}
}

@article{sorensen2021analysis,
  title={Analysis of residential EV energy flexibility potential based on real-world charging reports and smart meter data},
  author={S{\o}rensen, {\AA}se Lekang and Lindberg, Karen Byskov and Sartori, Igor and Andresen, Inger},
  journal={Energy and Buildings},
  volume={241},
  pages={110923},
  year={2021},
  publisher={Elsevier}
}

@report{iea2026electricity,
  author       = {{International Energy Agency}},
  title        = {Electricity Market Report 2023},
  year         = {2023},
  institution  = {IEA},
  url          = {https://www.iea.org/reports/electricity-2026},
}

@article{norris2025rethinking,
  title={Rethinking load growth: Assessing the potential for integration of large flexible loads in us power systems},
  author={Norris, Tyler and Profeta, Timothy and Patino-Echeverri, Dalia and Cowie-Haskell, Adam},
  year={2025},
  publisher={Nicholas Institute for Energy, Environment \& Sustainability}
}

@article{madavan2023risk,
  title={Risk-based hosting capacity analysis in distribution systems},
  author={Madavan, Avinash N and Dahlin, Nathan and Bose, Subhonmesh and Tong, Lang},
  journal={IEEE Transactions on Power Systems},
  volume={39},
  number={1},
  pages={355--365},
  year={2023},
  publisher={IEEE}
}

@article{yuan2025investigation,
  title={Investigation on Electricity Flexibility and Demand-Response Strategies for Grid-Interactive Buildings},
  author={Yuan, Haiyang and Chen, Yongbao and Chen, Zhe},
  journal={Buildings},
  volume={15},
  number={23},
  pages={4368},
  year={2025},
  publisher={MDPI}
}

@article{almutairi2024hierarchical,
  title={A hierarchical optimization approach to maximize hosting capacity for electric vehicles and renewable energy sources through demand response and transmission expansion planning},
  author={Almutairi, Sulaiman Z and Alharbi, Abdullah M and Ali, Ziad M and Refaat, Mohamed M and Aleem, Shady HE Abdel},
  journal={Scientific Reports},
  volume={14},
  number={1},
  pages={15765},
  year={2024},
  publisher={Nature Publishing Group UK London}
}

@article{ahmad2025demand,
  title={Demand response program towards sustainable power supply: current status, challenges, and prospects in Malaysia},
  author={Ahmad, Mohammad Sameer and Mansor, Nurulafiqah Nadzirah and Mokhlis, Hazlie and Naidu, Kanendra and Mohamad, Hasmaini and Ramadhani, Farah},
  journal={IEEE Access},
  year={2025},
  publisher={IEEE}
}

@article{yasmin2024survey,
  title={A survey of commercial and industrial demand response flexibility with energy storage systems and renewable energy},
  author={Yasmin, Roksana and Amin, BM Ruhul and Shah, Rakibuzzaman and Barton, Andrew},
  journal={Sustainability},
  volume={16},
  number={2},
  pages={731},
  year={2024},
  publisher={MDPI}
}

@ARTICLE{10577755,
  author={Zahari, Nur Elida Mohamad and Mokhlis, Hazlie and Mubarak, Hamza and Mansor, Nurulafiqah Nadzirah and Sulaima, Mohamad Fani and Ramasamy, Agileswari K. and Zulkapli, Mohd Faisal and Ja’Apar, Muhammad Asraf Bin and Jaafar, Mashitah and Marsadek, Marayati Binti},
  journal={IEEE Access}, 
  title={Integrating Solar PV, Battery Storage, and Demand Response for Industrial Peak Shaving: A Systematic Review on Strategy, Challenges and Case Study in Malaysian Food Manufacturing}, 
  year={2024},
  volume={12},
  number={},
  pages={106832-106856},
  doi={10.1109/ACCESS.2024.3420941}}

\end{document}